\documentclass[journal,12pt,onecolumn,draftclsnofoot,usenames,dvipsnames]{IEEEtran}

\usepackage{graphicx}
\usepackage{ifpdf}
\ifpdf
   \DeclareGraphicsExtensions{.pdf,.png,.jpg,.mps}
   \usepackage{url}
\else
\fi

\usepackage{mystyle}

\begin{document}

\title{Enabling Mobility in LTE-Compatible Mobile-edge Computing with Programmable Switches}
\author{\IEEEauthorblockN{
    Ashkan Aghdai\IEEEauthorrefmark{1},
    Yang Xu\IEEEauthorrefmark{2},
    Mark Huang\IEEEauthorrefmark{3},
    David H. Dai\IEEEauthorrefmark{3},
    H. Jonathan Chao\IEEEauthorrefmark{1}}
\IEEEauthorblockA{
    \\ \IEEEauthorrefmark{1}Tandon School of Engineering, New York University, Brooklyn, NY, USA
    }
\and
\IEEEauthorblockA{
    \\ \IEEEauthorrefmark{2}Department of Computer Science, Fudan University, Shanghai, China
    }
\and
\IEEEauthorblockA{
    \\ \IEEEauthorrefmark{3}Huawei Technologies, Santa Clara, CA, USA
}
}

\newcommand\copyrighttext{%
  \footnotesize This article is submitted to IEEE Journal on Selected Areas in Communications and is under review.}
\newcommand\copyrightnotice{%
\begin{tikzpicture}[remember picture,overlay]
\node[anchor=south,yshift=50pt] at (current page.south) {\fbox{\parbox{\dimexpr\textwidth-\fboxsep-\fboxrule\relax}{\copyrighttext}}};
\end{tikzpicture}%
}

\maketitle

\begin{abstract}
    Network softwarization triggered a new wave of innovation in modern network design.
    The next generation of mobile networks embraces this trend.
    Mobile-edge computing (MEC) is a key part of emerging mobile networks that enables ultra-low latency mission-critical application such as vehicle-to vehicle communication.
    MEC aims at bringing delay-sensitive applications closer to the radio access network to enable ultra-low latency for users and decrease the back-haul pressure on mobile service providers.
    However, there are no practical solutions to enable mobility at MEC where connections are no longer anchored to the core network and serving applications are supposed to move as their users move.
    We propose the mobile-edge gateway (MEGW) to address this gap.
    MEGW enables mobility for MEC applications transparently and without requiring any modifications to existing protocols and applications.
    MEGW supports mobility by reconstructing mobile users' location via listening to LTE control plane in addition to using two-stage location-dependent traffic steering for edge connections.
    Networks can incrementally upgrade to support MEC by upgrading some IP router to programmable switches that run MEGW.
    We have implemented MEGW using P4 language and verified its compatibility with existing LTE networks in a testbed running reference LTE protocol stack.
    Furthermore, using packet-level simulations we show that the two-stage traffic steering algorithm reduces the number of application migrations and simplifies service provisioning.
\end{abstract}

\copyrightnotice


\section{Introduction}

Software-defined networking (SDN) and network function virtualization (NFV) have revolutionized the design and implementation of modern networks.
Mobile networks are no exception to this trend as we see more and more components of these networks are softwarized.
The next generation of mobile networks (5G) relies on SDN and NFV as a foundation of its deployments.
5G shifts many paradigms in mobile networks, chief among which is the unification of various access technologies including 3GPP-compliant and non-compliant radio access networks to build a massive unified network with a huge capacity and ultra-low latency~\cite{5garch}.
The ultra-low latency promise of 5G opens new doors for development and deployment of the next generation of mobile applications such as vehicle-to-vehicle (V2X)~\cite{zheng2015heterogeneous}.
However, the low latency guarantees of 5G are not achievable with the tried and tested model of cloud computing since cloud-hosted services and applications sit at the core networks and have massive end-to-end latency to mobile devices due to the great physical distance.
Therefore, to build the next generation of mission-critical applications with ultra-low latency, another paradigm shift is required: from core cloud computing to edge cloud computing.
Mobile-edge computing (MEC) is a critical component 5G for achieving low latency~\cite{hu2015mobile,chen2016efficient} and realizing the shift.
Moving mission-critical applications to the edge cloud of mobile networks enables us to achieve ultra-low latency.

Implementing MEC is a challenging task when taking into account of supporting mobility among the MECs.
Before the introduction of MEC, the packet data network gateway (PGW) was an IP anchor point for all mobile connections~\cite{networkArch}.
As such, network state of connections did not need to move and was anchored at PGW.
However, consider the same scenario for a connection from a mobile user to an edge service in an MEC network.
If the user moves to a new location, the edge server may also need to move to a new MEC cluster that is closer to the new location for the user to maintain the low latency.
Also, if a similar service already exists in the new location of the user, it cannot handle users requests since the state of the user is held at the old location.
Therefore, it is necessary to migrate all or parts of the application that served the moving user from the old location to the new location.
In this case, the connection is no longer anchored at PGW or any other location at the network.
As such, a simple handover to a new radio base station may involve a lot of networking and application migration to a new location in the background.
This simple example shows that the emerging MEC technologies require flexibility and agility that is unprecedented in legacy networks and is only achievable through extensive use of programmability and softwariztion of network components.
We propose using modular software design patterns and the programmability of SDN to realize MEC in a 5G network without changing any of the components from devices to servers, applications, and network protocol stack.

Recent studies~\cite{li2018mobile,M-CORD,MAGMA} including our own previous work~\cite{aghdai2018transparent} demonstrate the feasibility of implementing MEC and steering edge-bound traffic toward it.
However, none of the previous studies tackle the challenges that are involved with mobile users.
In this paper, we design and implement the mobile-edge gateway (MEGW) using programmable switches.
Existing mobile networks can offer MEC applications by upgrading some of the IP routers in the radio access network (RAN) to programmable switches running MEG.
MEGW is transparent to existing components of emerging 5G networks as well as the network protocol stack of mobile users' devices and servers that host mobile services; therefore, existing network components require no modifications when offering MEC edge applications using MEGW.
The MEGW is programmed entirely in P4 language, is ported to a Netronome NFP4000~\cite{nfp4000} P4 target, and is verified to work with LTE protocol stack in a testbed running OpenAirInterface~\cite{nikaein2014openairinterface} platform.
Using the MEGW, we have implemented content delivery at the mobile edge to reduce latency for users and back-haul pressure for operators.
We expect MEC services to follow the scale-out design pattern by deploying many instances of services based on users' demand.
As a result, efficient load balancing will play a critical role in the overall performance of MEC applications.
Therefore, we believe that load balancing among instances of an MEC application should be included in the MEGW, and have implemented this function in our prototype.

MEGW is a practical solution for deploying MEC services on existing LTE networks and emerging 5G networks.
We make the following contributions.
\begin{itemize}
    \item Design, implementation, and testbed verification of MEGW steering traffic to an edge service using OpenAirInterface~\cite{nikaein2014openairinterface} reference 5G protocol stack and fully operational control and data planes of MEGW running on a programmable switch.
    \item A 2-stage algorithm for steering edge traffic to applications aiming at minimizing the number of application state migrations to enable scalability.
    \item Simplifying the provisioning of MEC resources based on the hierarchy of 5G networks.
    \item Proposing an end-to-end solution for establishing ultra-low latency connections to MEC without any modifications to reference 5G protocol stack, client/servers protocol stack, or existing components of 5G networks.
        Our solution also does not involve additional LTE control plane operations as it does not terminate GTP tunnels.
        Moreover, it only requires the gradual upgrade of \emph{some (and not all) of} legacy IP routers to programmable switches.
\end{itemize}

The rest of the paper is organized as follows.
\S~\ref{section:motivation} reviews the architecture of LTE networks, our previous work on MEC traffic steering, and lays out the mobility problem.
\S~\ref{section:mobility} presents a two-step traffic steering solution to minimize application migrations at edge.
\S~\ref{section:MEGW} introduces MEGW, highlights its differences with existing works, and proposes a practical solution for providing mobility for MEC connections.
\S~\ref{section:results} verifies the operation of MEGW using reference 5G protocol stack and evaluates its performance using various testbed and simulation experiments.
\S~\ref{section:related} reviews the related works.
Finally, \S~\ref{section:conclusion} concludes the paper.

\section{Motivation and Background}\label{section:motivation}

\begin{figure*}[t]
    \centering
    \includegraphics[width=1\textwidth]{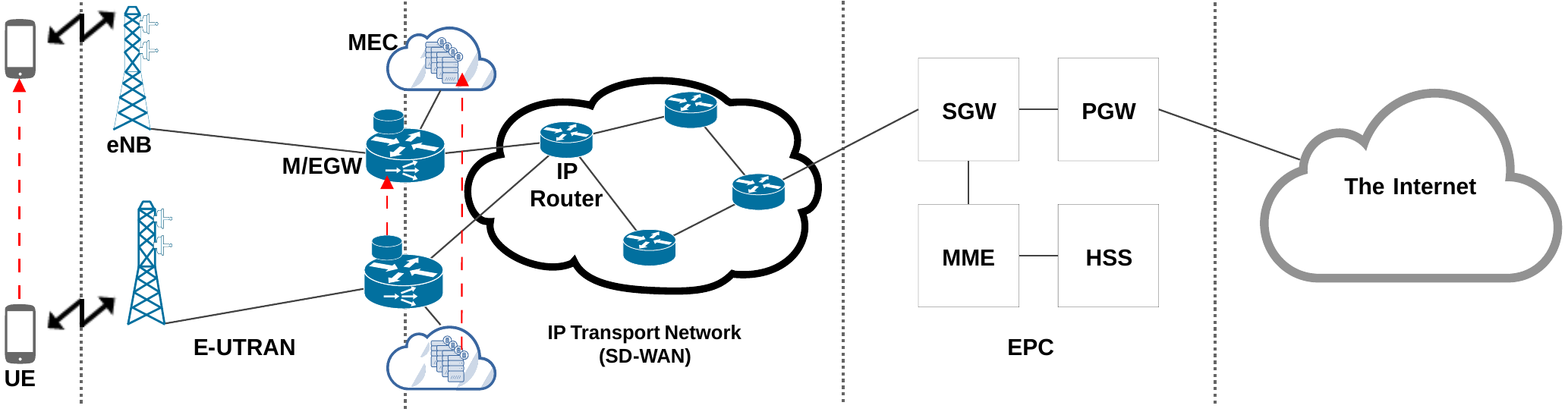}
    \caption{Mobility complicates traffic steering to the mobile-edge cloud}
    \label{fig:overview}
\end{figure*}

Currently, cloud computing is the dominant design pattern for implementing web applications.
In this architecture, huge warehouse-scale data centers located at the core of the Internet host web and mobile applications and services alike.
While hosting applications at the core cloud enables scalability, elasticity, ease of maintenance, and high performance for mobile applications, it has a significant disadvantage in adding a considerable amount of delay.
Some mobile applications may tolerate the added latency. However, the vast majority of mission-critical mobile applications require low latency, and thus cannot be hosted at core clouds.
V2X is an excellent example of such applications.
With the introduction of self-driving cars and their reliance on V2X, an increase in end-to-end latency could lead to degraded coordination among self-driving vehicles that could easily result in a tragedy.

5G mobile networks aim to integrate different access technologies into an unified network with massive capacity and low latency.
MEC is a significant component of 5G for achieving high throughput and low latency.
Figure~\ref{fig:overview} illustrates the architecture of a 5G network~\cite{networkArch} with the addition of MEC.
The user equipment (UE), the communication device used by the users, connects to the RAN (evolved universal terrestrial radio access network or E-UTRAN in LTE~\cite{eutran}) through base stations (evolved Node B or eNB in LTE).
Mobile core network (evolved packet core or EPC) includes the packet data network gateway (PGW) as the point of contact with packet networks like the Internet, the serving gateway (SGW) that acts as a router between eNBs and the PGW, mobility management entities (MMEs) that control the operation of mobile devices and their signaling, and home subscriber service (HSS) that acts as a database containing subscribers' information.
The EPC typically includes multiple SGWs for different geographical regions that forward packets between the PGW and their eNBs.
MEC clusters are located in close proximity of the radio edge of the mobile network to shorten connections' datapath and enable ultra-low latency for mission-critical applications such as V2X.
In our previous work, we introduced the edge gateway (EGW)~\cite{aghdai2018transparent}, a programmable L3 switch capable of transparently steering edge traffic back and forth to MEC clusters.

Mobility is the essential function of mobile networks as it enables the users to move freely in the areas covered by the RAN and use voice/data services seamlessly without any disruption.
However, neither our previous work~\cite{aghdai2018transparent} nor the rest of the studies in this area~\cite{li2018mobile,M-CORD,MAGMA} consider the ramifications of users' mobility.
Prior to the introduction of MEC, in 4G LTE mobile networks all voice and data connections are anchored at the EPC.
Meaning that after UEs perform a \emph{handover} between two eNBs, packets' point of contact with the Internet does not change.
As such, in the handover procedure, only the network state of UEs radio bearers is transferred from the old eNB to the new eNB.
However, as shown in Figure~\ref{fig:overview}, handovers that involve edge connections are more complicated than the LTE handover since edge connections are no longer anchored at PGW.
During the handover, edge connections' destination moves with the device to an MEC closest to the new eNB.
Therefore, a 5G handover involving MEC connections include two types of migrations:
\begin{enumerate}
    \item \textbf{Type I: Migration of network state.} This procedure is similar to a typical LTE handover where radio bearers information is transferred from an old eNB to a new eNB and GTP tunnels are modified accordingly to update UEs datapath to/from EPC.
    Additionally, the GTP context of EGW(s) connected to involved eNBs should be updated to update UEs' datapath to the serving MEC.
    \item \textbf{Type II: Migration of edge applications' state.} In cases where the serving application is to be moved from the old MEC to the new MEC. 
    A naive solution for this problem is to move the whole serving instance (virtual machine or application container) from an old MEC cluster to a new cluster which is extremely costly considering the high frequency of UE handovers in a production mobile network.
    Machen et al.~\cite{machen2018live} propose a layered backend design for edge services to minimize the cost of this type of migrations and avoid complete VM migrations.
\end{enumerate}

We focus on mobility with presence of MEC by extending EGW to support handover in various operating scenarios.
We aim at building a scalable low-latency infrastructure for MEC with aid of network softwarization by meeting the following objectives:
\begin{enumerate}
    \item Type I migrations should be performed transparently and without additional control plane involvement to enable scalability at the control plane and low latency at the data plane.
    \item The frequency of Type II migrations should be minimized to achieve scalability as they involve large data transfers and potential down times. At the same time, Type II migrations enable low-latency connections to the edge. We aim to optimize the network infrastructure for MEC by designing a low-latency system with a reduced number of application state migrations.
\end{enumerate}

Next, we review the basics of LTE networks, MEC traffic steering, the architecture of EGW, and the handover procedure according to 4G LTE reference.

\subsection{Components of LTE Networks}\label{sebsection:LTE}

In LTE networks, the RAN communicates with the EPC by using the S1 interface.
This interface defines the GTPv1-U~\cite{gtpu} protocol as the data plane that carries user traffic between eNB and SGW.
S1 interface also defines signaling messages - or the control plane - between eNBs and MME for controlling and locating UEs.
The S1 application protocol (S1AP)~\cite{s1ap} is used for the control plane.

MMEs assign hop-by-hop tunnel endpoint identifiers (TEID) between UE and PGW once UEs attach to the network;
TEIDs on upstream (eNB-to-SGW) and downstream (SGW-to-eNB) data paths are assigned through S1AP protocol.
TEID is used for identification and routing of data plane GTP-U tunnels.
GTP-U encapsulates the IP packets from the UE in a data plane tunnel; it includes an outer IPv4 header, a UDP header, and the GTPv1-U header including TEID.
IP routers forward packets between the RAN and the EPC using only the outer IPv4 headers.
As a result of using GTP-U tunnels, network operators could not deploy network functions such as content delivery in the IP transport network; therefore, network functions/services are deployed at PGW that terminates the GTP-U tunnel or the IPX that connects PGW to the Internet.
This is a source of additional delay for mobile users as well as additional traffic at the core network for the operators.

\subsection{Edge Gateway Architecture}\label{subsection:EGW}

Consider the mobile network of Figure~\ref{fig:EGW:arch}.
EGW allows mobile users use applications/services hosted by MEC clusters.
To achieve that, EGW relies on the origin server of the application to redirect the requests from UE1 to the virtual IP address (VIP) of the corresponding MEC application.
The EGW will take over the rest of the communication: Once UE1 sends a request to the VIP as an inner IP destination, EGW steers the traffic to one of the serving instances of the MEC application that is configured with a direct IP address (DIP).

The EGW is implemented as three add-on modules on top of an IP router. They are Service Offloader, Load Balancer, and S1AP Processor. 
The first two modules provide extra data plane functions, while the third module acts as a controller that reconstructs the state of GTP tunnels by listening to S1AP communications between eNBs and MME.
These three modules are briefly described below:
\begin{figure*}[t]
    \centering
    \includegraphics[width=1\textwidth]{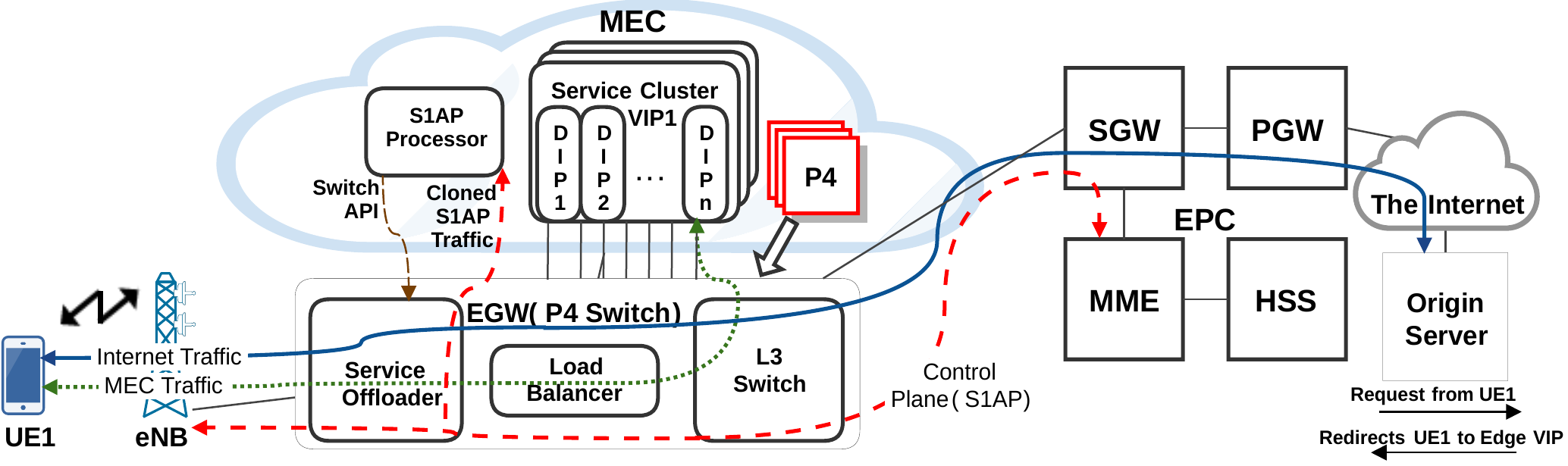}
    \caption{Softwarized implementation of EGW with aid of P4 programming language.}
    \label{fig:EGW:arch}
\end{figure*}

\subsubsection{Service Offloader}
This module handles S1 interface packets between eNB and SGW.
It offloads MEC-bound GTP packets to a local service cluster and repackages IP packets from the cluster back to GTP seamlessly.

Service offloader implements the following primary functions of the EGW:
\begin{itemize}
    \item S1AP control messages are cloned to an interface connected to the S1AP processor.
        The original packet is sent to the IP router unmodified and will get forwarded to the correct destination according to its outer destination IP address.
    \item Upstream GTP-U packets from eNB are matched against their inner destination IP address.
        For VIP-bound packets, the inner IP header will be rewritten to the outer IP header, while their GTP protocol stack (GTP-U, UDP, and inner IP headers) is removed.
        As a result, the following modules will use the inner IP header to forward packets to VIP.
    \item Downstream IP packets from a DIP in a service cluster to a UE are encapsulated in a GTP-U tunnel.
        More specifically, the original IP header of these packets is copied to the inner IP header, while an outer IP header with eNB and SGW IP addresses, a UDP header with GTP-U port designator, and a GTP-U header with UE's corresponding downstream TEID are added. 
        This module requires GTP-U context which is a mapping between UEs' IP address and their downstream TEID.
        S1AP processor module reconstructs this context as forwarding rules for the service offloader.
\end{itemize}

\begin{figure}[t]
    \centering
    \includegraphics[width=0.95\linewidth]{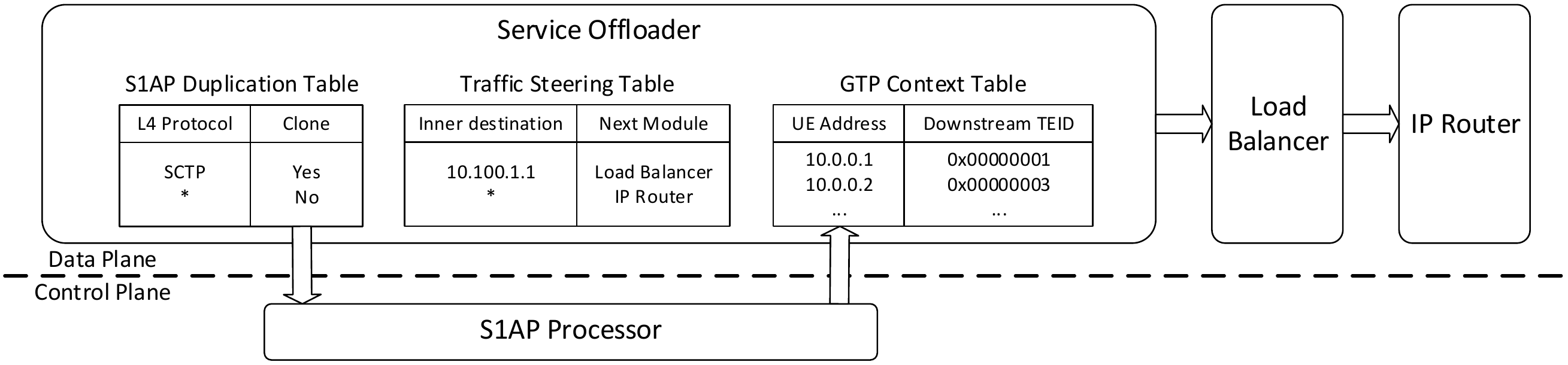}
    \caption{Implementation design of EGW control and data planes.}
    \label{fig:EGW:p4}
\end{figure}

\subsubsection{Load Balancer}
The primary task of this module is to choose a serving instance for packets destined to VIPs, i.e., rewrite DIP in place of VIP if the outer destination IP address is that of a VIP.

\subsubsection{S1AP Processor}
This module receives cloned S1AP packets and parses them according to ASN.1~\cite{steedman1993abstract} and Non-Access-Stratum (NAS)~\cite{NAS} specification provided by the LTE standard.
The S1AP processor reconstructs the UE to downstream TEID mapping by listening to two specific S1AP messages between eNB and MME that are sent once UE attaches to the eNB: \emph{InitialContextSetupRequest} and \emph{InitialContextSetupResponse} messages.
The former assigns the upstream TEID (eNB-to-SGW data path), while the latter assigns the downstream TEID (SGW-to-eNB data path) for the attached UE.
Once the S1AP processor receives both messages for an attached UE, it will call the switch API to add a rule that maps UE's IP address to corresponding downstream TEID at the service offloader module.

This solution implicitly assumes that each UE is only assigned a single Bearer to the PGW.
We omit this assumption in the design of MEGW and offer a multi-bearer solution in \S~\ref{subsection:megw:arch}.

\subsection{P4 Implementation of EGW data plane}
Figure~\ref{fig:EGW:p4} demonstrates the detail design of the three implemented data plane modules using P4 language. As indicated in the figure, only the service offloader module interfaces with the S1AP processor.

The service offloader classifies packets based on their protocol.
The S1AP duplication table clones matched packets with a valid Stream Control Transmission Protocol (SCTP) header to the S1AP processor, and the original is sent to the IP router.
Traffic steering table matches on valid GTP-U header and the inner destination IP address and sends matched packets to the load balancer with the GTP protocol stack removed. The figure assumes a VIP with address 10.100.1.1.
GTP-U context table contains the UE to downstream TEID (32-bit identifier for GTP tunnels) mapping; it matches on the destination IP address of downstream edge connections (which is UE's IP address), encapsulates UE-bound packets in their corresponding GTP-U tunnel, and sends them to the IP router.

The load balancer module applies to VIP-bound packets and redirects them to DIPs.
The connection-to-DIP mapping should remain consistent during the lifetime of connections.
Any layer 4 load balancer that meets the consistency requirement can be used here.
Silkroad~\cite{miao2017silkroad}, Beamer~\cite{olteanu2018stateless}. and Spotlight~\cite{aghdai2018spotlight} are great candidates as they are portable to programmable switches as a load balancing module with the consistency guarantee.

Service offloader and load balancer modules pass on the incoming packets to a standard IP router module, which performs all the necessary routing functions similar to a legacy IP router used in the IP transport network.

\subsection{UE handover procedure}\label{subsection:background:handover}

\begin{figure}[t]
    \centering
    \includegraphics[width=0.6\linewidth]{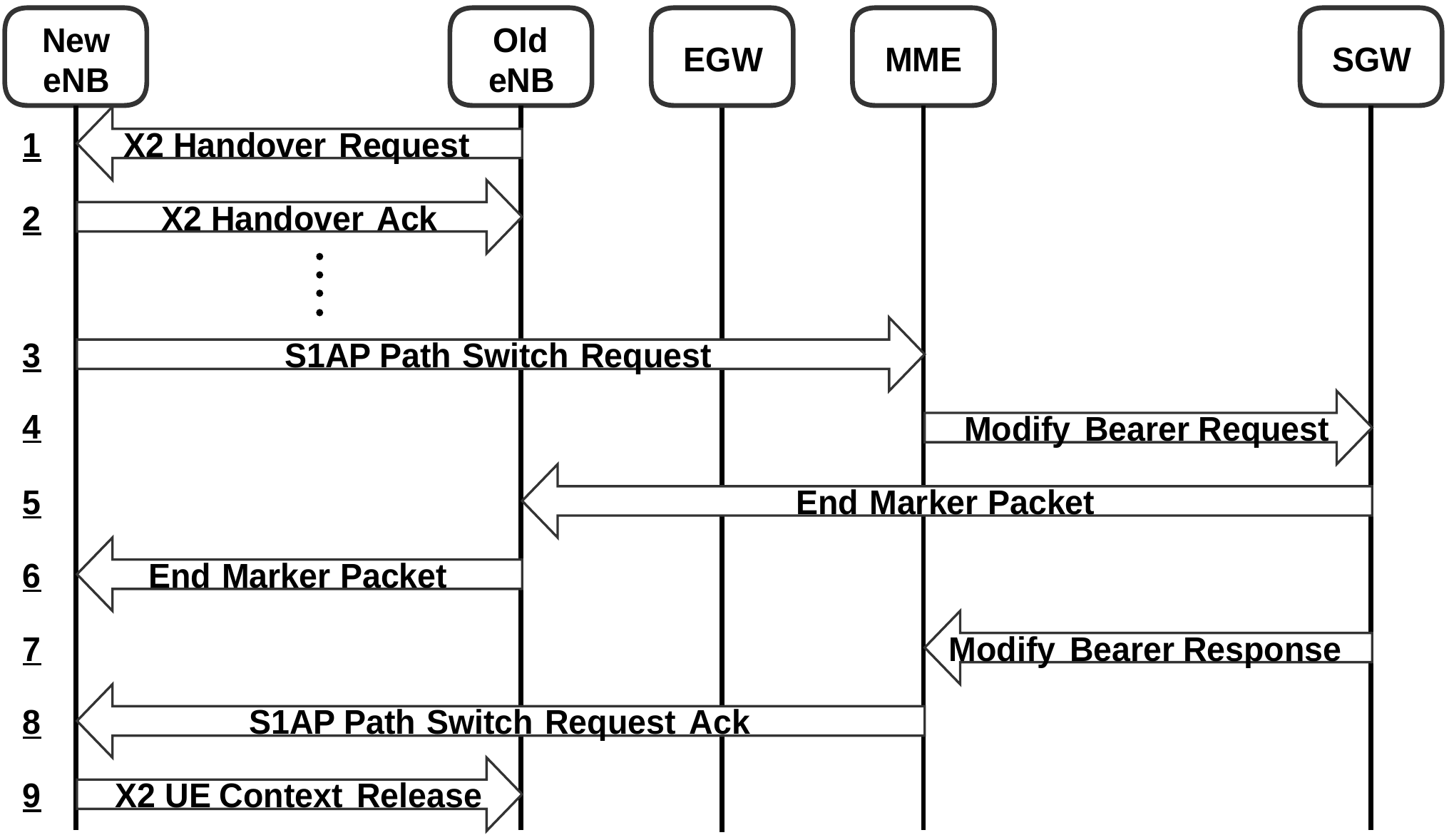}
    \caption{Timeline of an X2-based handover procedure~\cite{lteho1,lteho2}.}
    \label{fig:LTE:HO}
\end{figure}

Figure~\ref{fig:LTE:HO} shows a subset of LTE control plane messages transferred during an X2 handover.
X2 handovers take place when UEs identify that a neighboring cell has a better signal quality in their signal quality measurements.
The EGW does not terminate X2 and S1AP signaling involved in this procedure.
However, we added the EGW to the figure to highlight which messages pass through it.
Later we use these messages to identify the source and destination eNBs in X2 handovers transparently.

The procedure starts when the old eNB sends an X2 request to the new eNB as a result of UE signal quality measurements.
X2 packets are not routed through the EPC; hence these messages do not traverse through EGW.
Upon acknowledging the handover, the new eNB sends a path switch request to MME so that MME recognize the new location of the UE and establish GTP tunnels between SGW and the eNB to complete the handover.
After the handover is acknowledged by the new eNB (step 2) the UE attaches to the new eNB.
However, before the path switch message is acknowledged by the MME (step 8), new eNB forwards the upstream traffic to the old eNb.
However, since no GTP tunnel is established for UE packets between the new eNB and SGW, the new eNB forwards the packets to/from the UE to old eNB using X2 interface.
EGW observes the S1AP path switch request (step 3) and forwards it to the S1AP processor. Upon processing these messages, we can identify the start of X2 handover and the IP addresses of old and new eNBs.
Once the MME receives the path switch request, it modifies the path for all bearers of the UE and informs the SGW using modify bearer request.
Modify bearer request is routed inside the EPC and is not visible to the EGW.
SGW terminates the GTP tunnels to old eNB for UE bearers and marks the end of communications to the old eNB through the end marker packet.
End marker packet is the last data plane packet routed from the SGW to old eNB path and is observed by EGW (step 6).
End marker packet marks the end of the forwarding phase using the X2 interface between eNBs.
Finally, the SGW sends an acknowledgment to MME about the new GTP tunnels to new eNB and only then does the MME acknowledge the S1AP path switch request to the new eNB.
The S1AP acknowledgment is also observed by EGW (step 8) and marks the completion of the X2 handover process.
After step 5 and before step 7, SGW is not supposed to send any packets to the UE.
The LTE reference does not specify if downstream packets queued at SGW destined to the moving UE should be buffered or dropped during this duration and leaves the implementation details to device vendors.
We refer to this period as the silence period and note that it is extremely short -- an RTT between MME and SGW in the worse case.
Edge service should also be silent in this period and do not send any packets to the old or new eNB after step 5 and before step 8. 
During the forwarding phase -- before step 8 -- EGW should send downstream packets to the old eNB using the old GTP tunnels.
Downstream packets are forwarded to the new eNB using the newly established GTP tunnels after step 8.

\section{Mobility at Edge}\label{section:mobility}

\begin{figure}[t]
    \centering
    \includegraphics[width=0.6\textwidth]{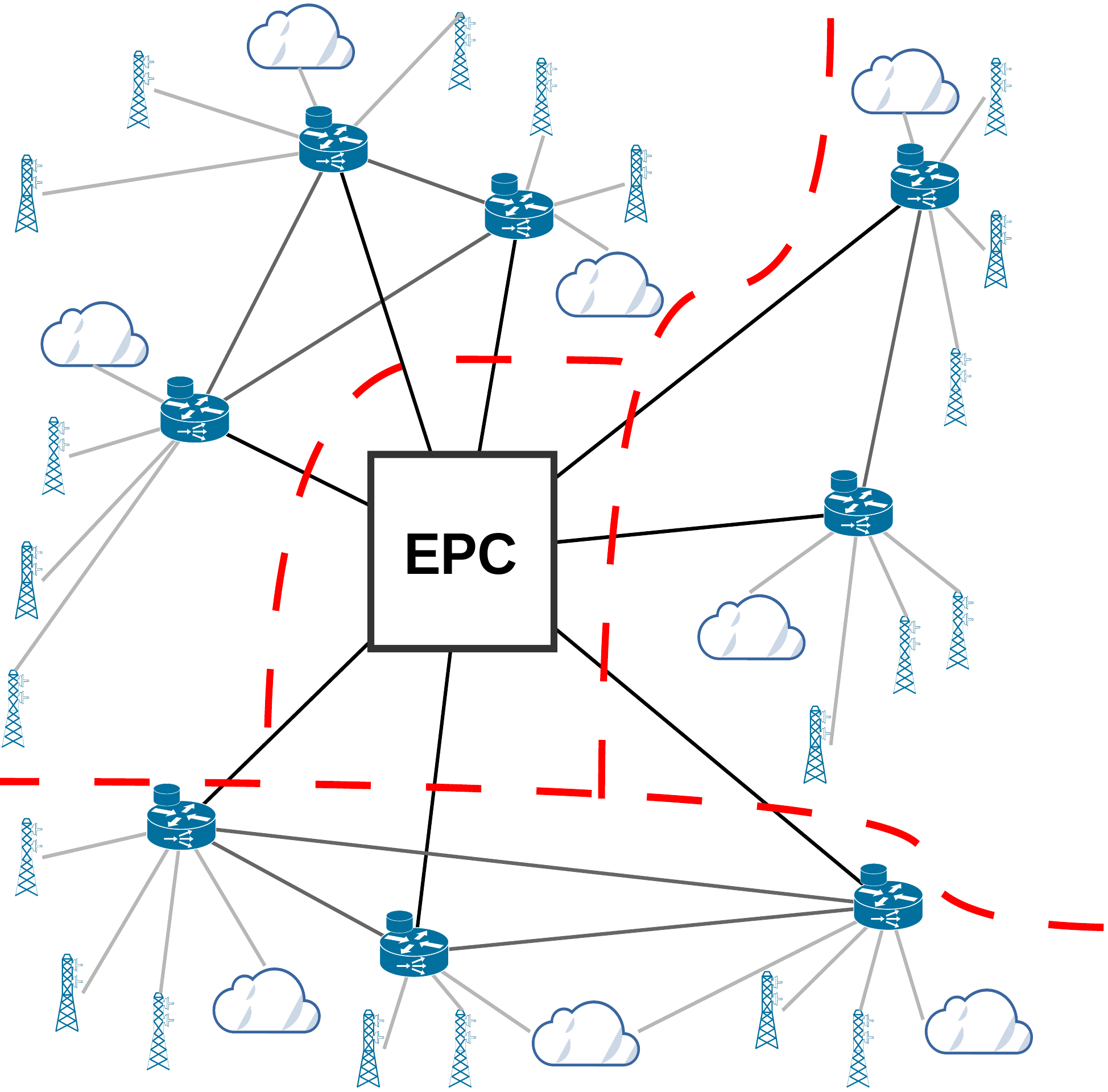}
    \caption{Hierarchy of 5G networks allows us to define smaller regions where cluster-to-cluster latency is smaller than radio latency.}
    \label{fig:5G:regions}
\end{figure}

Consider type II migrations, i.e., migration of serving applications' state to a new MEC when UEs move.
Such migrations are extremely expensive as they include service down time and huge traffic ovearhed to move parts or all of VMs or application containers~\cite{machen2018live}.
Figure~\ref{fig:5G:regions} illustrates the hierarchy of 5G networks as we introduce MEC.
We have partitioned the radio edge of the network into smaller regions identified by thick dashed lines in the figure.
The regions correspond to groups of MECs with close proximity in the same metropolitan area.
MEC clusters are typically deployed in telco central offices or as external rack containers close to eNBs.
Further, these clusters are typically connected using high-capacity software-defined WAN (SD-WAN)~\cite{att2017aec,att2016ecomp} to allow for aggregation of many small clusters into fewer large clusters for ease of management and usability using cloud operating systems such as OpenStack.
With the widespread usage of SD-WAN to connect MEC clusters we expect to observe very low latency for intre-cluster communications.
We define the regions in our hierarchical 5G networks as groups of MEC clusters such that the inter-cluster latency is smaller than the radio latency of 5G reference.
Given our assumptions, we propose to treat MECs in the same region as a single large cluster.
In other words, within a region, UEs' edge connections may be routed to any of available MEC clusters given that the edge service is available at all clusters.
As such, when UEs' move to a new eNB in the same region we do not migrate the serving application instance.
Instead, application state migrations only take place when UEs move between regions.

The above-proposed traffic steering can be implemented using a two-stage routing algorithm:
\begin{itemize}
    \item \textbf{Stage I:} Source MEGW (closest MEGW to UE) chooses a serving MEC in the current region using consistent hashing and route the edge traffic to the serving MEGW (closest MEGW to the serving MEC).
    \item \textbf{Stage II:} Serving MEGW uses a load balancing mechanism to steer edge connections to an instance of the edge service if more than one instance exists in the chosen MEC. Otherwise, packets are routed to the unique serving instance.
\end{itemize}

We show this process in Figure~\ref{fig:2stage:ho} where the UE is moved between two radios in the same region.
The source MEGWs choose the target MEGW using a consistent hashing mechanism, while the target MEGW performs load balancing for the virtual IP address (VIP) of the service if more than one instance exists.
The proposed architecture has many benefits including:

\begin{figure}[t]
    \centering
    \includegraphics[width=0.6\textwidth]{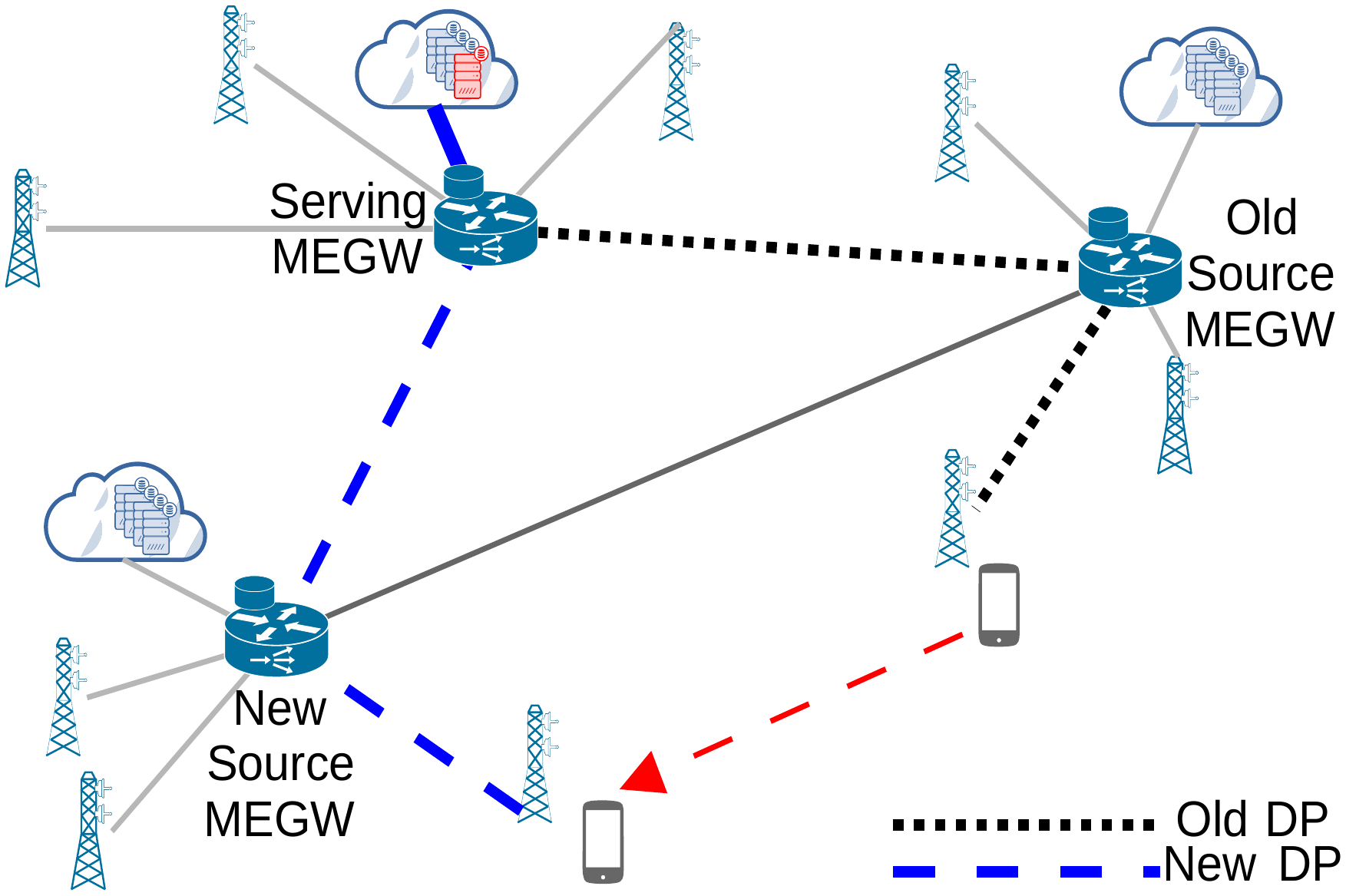}
    \caption{Two-stage traffic steering when UEs move within a region}
    \label{fig:2stage:ho}
\end{figure}

\begin{enumerate}
    \item The two-Stage MEGW traffic steering minimizes the frequency of application state migrations.
        Application state migration is an expensive process. Even state-of-the-art proposals~\cite{machen2018live} include service down times and hundreds of Megabytes of transfers per each application state migration.
        It is almost impossible to deploy edge services at scale if edge applications migrate for each UE handover in the system.
        As an example consider urban transit systems that move large masses of mobile users between train stations at high speed.
        The number of application migrations that these systems entail would be very large if application state is migrated for each UE handover.
        Introducing the regions heavily reduces the number of such handovers.
    \item The two-stage algorithm effectively creates a single massive logical cluster per region instead of many smaller physical clusters.
        Enlarging the clusters eases requirements of ISPs in terms of service provisioning and maintenance.
    \item The increased latency as a result of added hop is limited because in defining the boundaries of regions we limit the worse case maximum latency between any two MEGWs in the same region.
    \item Minimizing application state migrations decreases the amount of traffic due to migrations considerably.
\end{enumerate}

\section{Mobile Edge Gateway Architecture}\label{section:MEGW}

\begin{figure*}[t]
    \centering
    \includegraphics[width=1\textwidth]{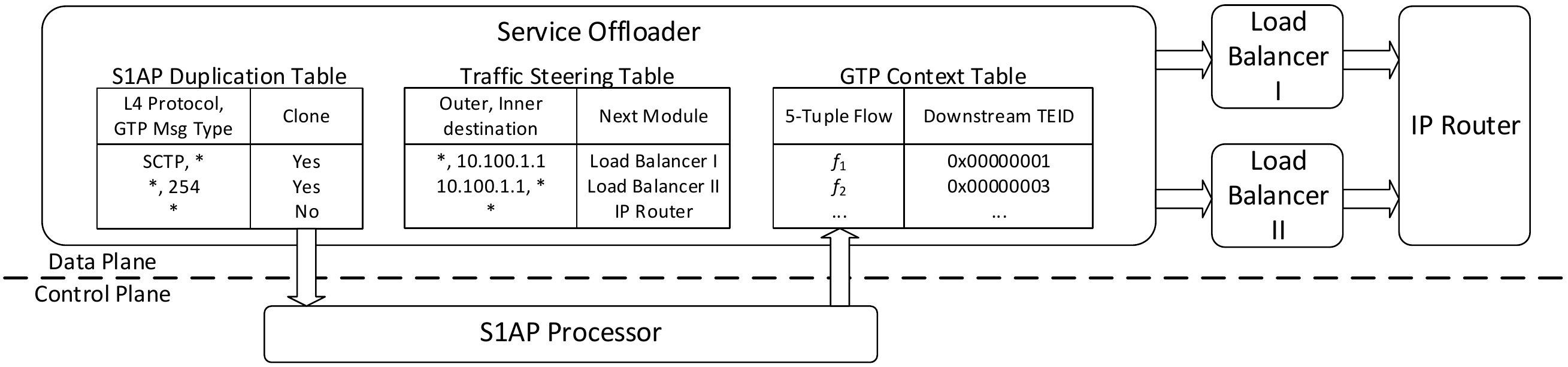}
    \caption{MEGW data plane forwarding components and their interactions with the control plane.}
    \label{fig:MEGW:arch}
\end{figure*}

MEGW is the evolution of EGW with added support for mobility and various optimizations to improve the scalability of its control plane.
As shown in Figure~\ref{fig:MEGW:arch}, the design of MEGW is modular.
All of the functions are well-supported by the existing programmable switches, and MEGW's data plane is implemented using P4 programming language.
MEGW's data plane is comprised of three modules (Service Offloader, Load Blancer I, and Load Balancer II) on top of an IP router.
MEGW's control plane relies on cloned S1AP attach (initial context setup request) and handover (path switch request) messages to reconstruct the LTE context (UE bearer to TEID mapping) when UEs attach to, detach from, or move between eNBs.

\subsection{MEGW Data Palne} \label{subsection:megw:arch}

MEGW data plane is evolved from EGW's architecture.
It builds a forwarding plane using add-on modules on top of an IP router.
MEGW's service offloader is entirely different from that of EGW's.
The modified service offloader clones S1AP packets as well as end marker packets to the controller (S1AP processor).
As indicated in the Figure, the S1AP duplication table is modified to recognize end marker packets using a field in the GTP header.
For end marker packets, the GTP message type is 254~\cite{gtpu}.

Furthernmore, we have optimized the traffic steering mechanism. 
While EGW matches on inner IP packets' destination and steers edge VIP-bound packets to the MEC cluster, MEGW matches on 5-tuple inner packet information (inner IP source and destination, L4 protocol type, source, and destination), meaning that the GTP context table is modified to match 5-tuple flows to UEs rather than UEs' IP address.
Upstream packets headers are cloned to the controller if and only if they miss the 5-tuple match.
The controller extracts the upstream bearer TEID, in addition to the 5-tuple flow information and installs two rules: an upstream rule with 5-tuple match so that the data plane does not clone packets from the same flow in the future and a downstream rule with corresponding downstream bearer TEID, so that edge packets are sent on the same bearer as upstream.
The new offloading mechanisem has two advantages:
\begin{enumerate}
    \item It is more scalable as rules are only installed for UEs that open edge connections.
        EGW, on the other hand, installs GTP context rules for \emph{all} UEs that attach to eNBs even if they do not open edge connections.
        Installing fewer rules on MEGW allows us to reduce the size of the table as well as the amount of rules installed by the controller. The reduced controller channel pressure improves the scalability of MEGW.
    \item By mapping the corresponding TEIDs to 5-tuple flows MEGW solves the multi-bearer problem.
        If a UE opens multiple edge connections on different bearers, each connection will be mapped to its corresponding bearer in the downstream rule.
\end{enumerate}

The first stage of load balancing (load balancer I in the figure) is applied to VIP-Bound packets with at the serving gateway.
These packets are encapsulated in GTP.
Traffic steering table recognizes such packets by matching on the destination address of inner IP header.
In the figure, we assumed the IP address of 10.100.1.1 for the VIP.
After decapsulating the packet, load balancer I uses consistent hashing to map edge connections to the serving EGW statelessly.
Meaning that if UEs move to a new eNB and EGW, the new EGW will apply the same stateless consistent hash on inner packets' headers and sends them to the same serving EGW.
The serving MEGW receives the decapsulated IP packets and the traffic steering table recognizes that the packets' destination IP address is VIP (10.100.1.1) and that they are not encapsulated in GTP.
As shown in the Figure, these packets are forward to load balancer II or the second stage of load balancing.
The second stage load balancer is applied to IP packets at the serving EGW.
It rewrites a DIP in place of the destination VIP and sends the packet to the IP router for routing to the serving instance of the service with the chosen VIP.
This stage of the load balancer is unaware of GTP protocol and location of UEs.
There is also a corner case where source MEGW happens to be the same as the serving MEGW (i.e., load balancer I hashes the packet to the same MEGW that first received it).
Load balancer I contains a branching table after the consistent hashing to recognize these packets and send them to load balancer II instead of the IP router.

\subsection{MEGW Control Plane}

Each MEGW includes an S1AP processor which acts as MEGWs' local controller.\footnote{We use the terms S1AP processor and MEGW controller interchangeably.}
S1AP processors' state is completely local to their corresponding MEGW, and the controllers do not have any shared state; neither do they require any central coordination.

S1AP processor parses cloned S1AP packets; for each UE the S1AP processor reconstructs a mapping from UE bearers to the corresponding upstream and downstream TEID pair.
When a UE attaches to an eNB, it will be assigned bearers and corresponding GTP tunnels for each bearer between eNB and SGW. S1AP processor observes S1AP attach request messages and responses and records the TEID pairs.
Similarly, when UEs move between two eNBs, the old eNB will send an S1AP path switch request message which is intercepted by the corresponding MEGW's S1AP processor and releases the old mapping.
The new eNB will receive an S1AP path switch acknowledgment which is observed by the new EGWs' S1AP processor and enables it to update the TEID pairs for all UE bearers.

\begin{figure*}[t]
    \centering
    \begin{minipage}{0.32\textwidth}
        \includegraphics[width=\linewidth]{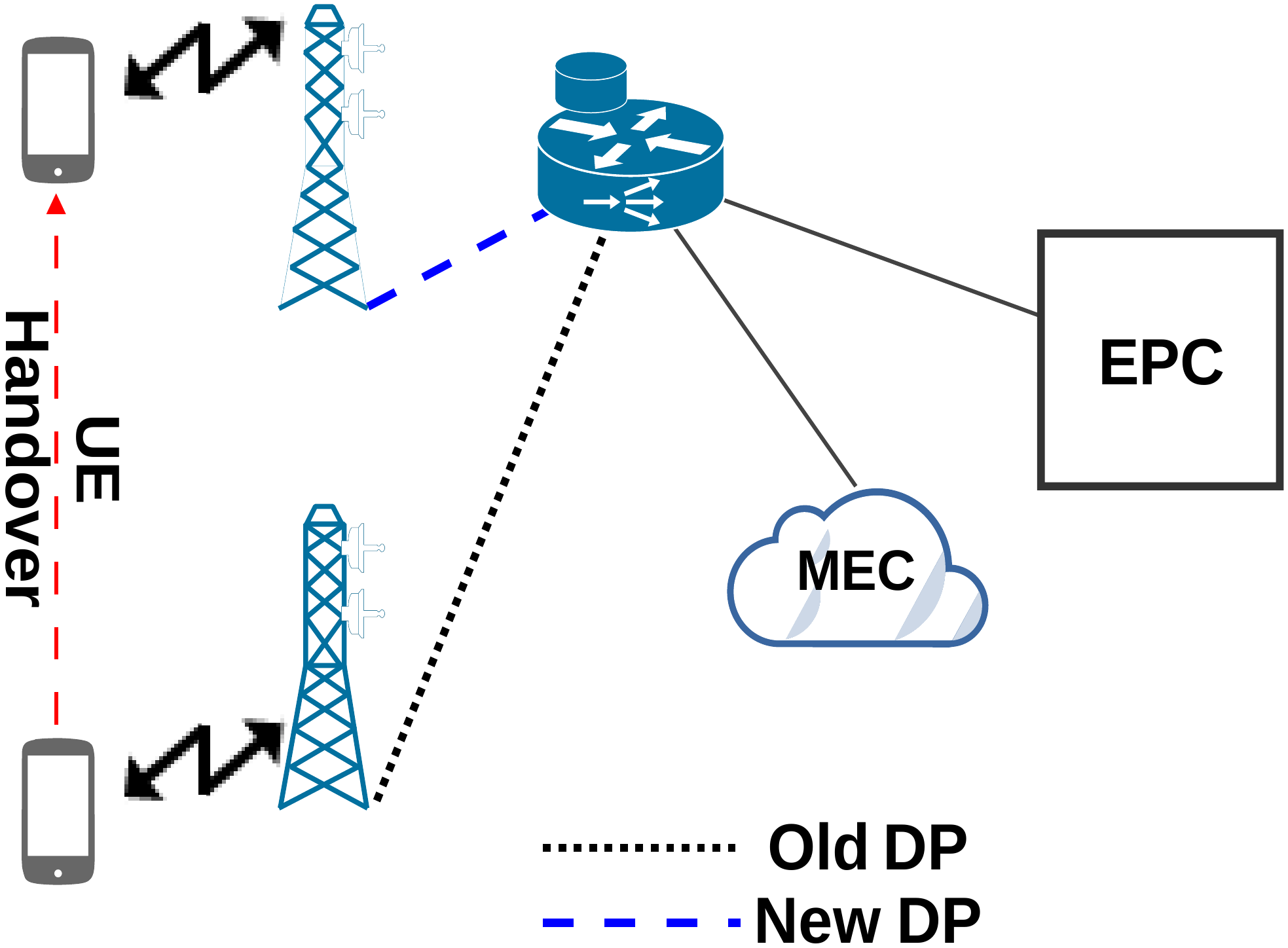}
        \label{migrationScenario1}
    \end{minipage}
	\begin{minipage}{0.32\textwidth}
		\includegraphics[width=\linewidth]{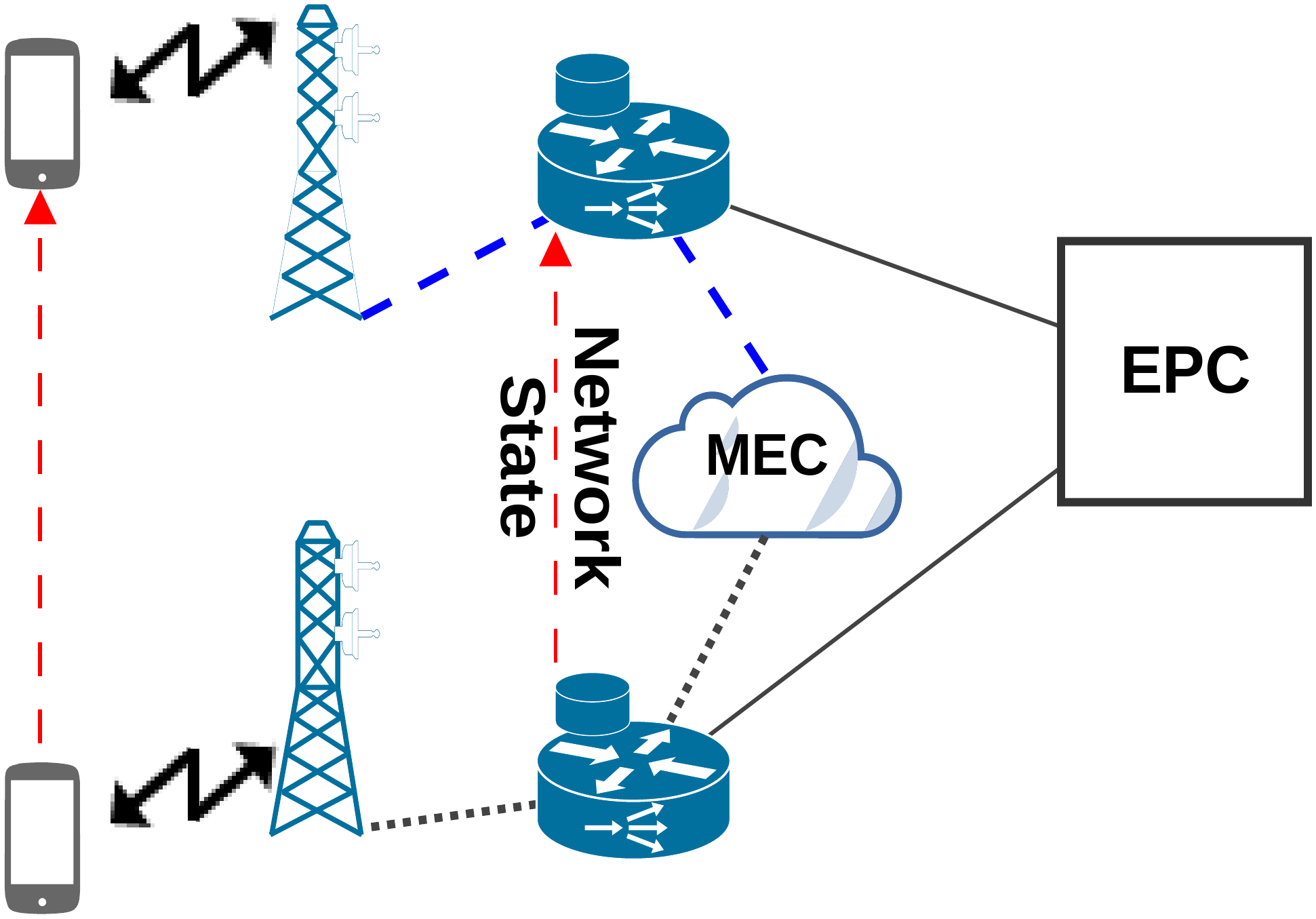}
		\label{migrationScenario2}
	\end{minipage}
	\begin{minipage}{0.32\textwidth}
		\includegraphics[width=\linewidth]{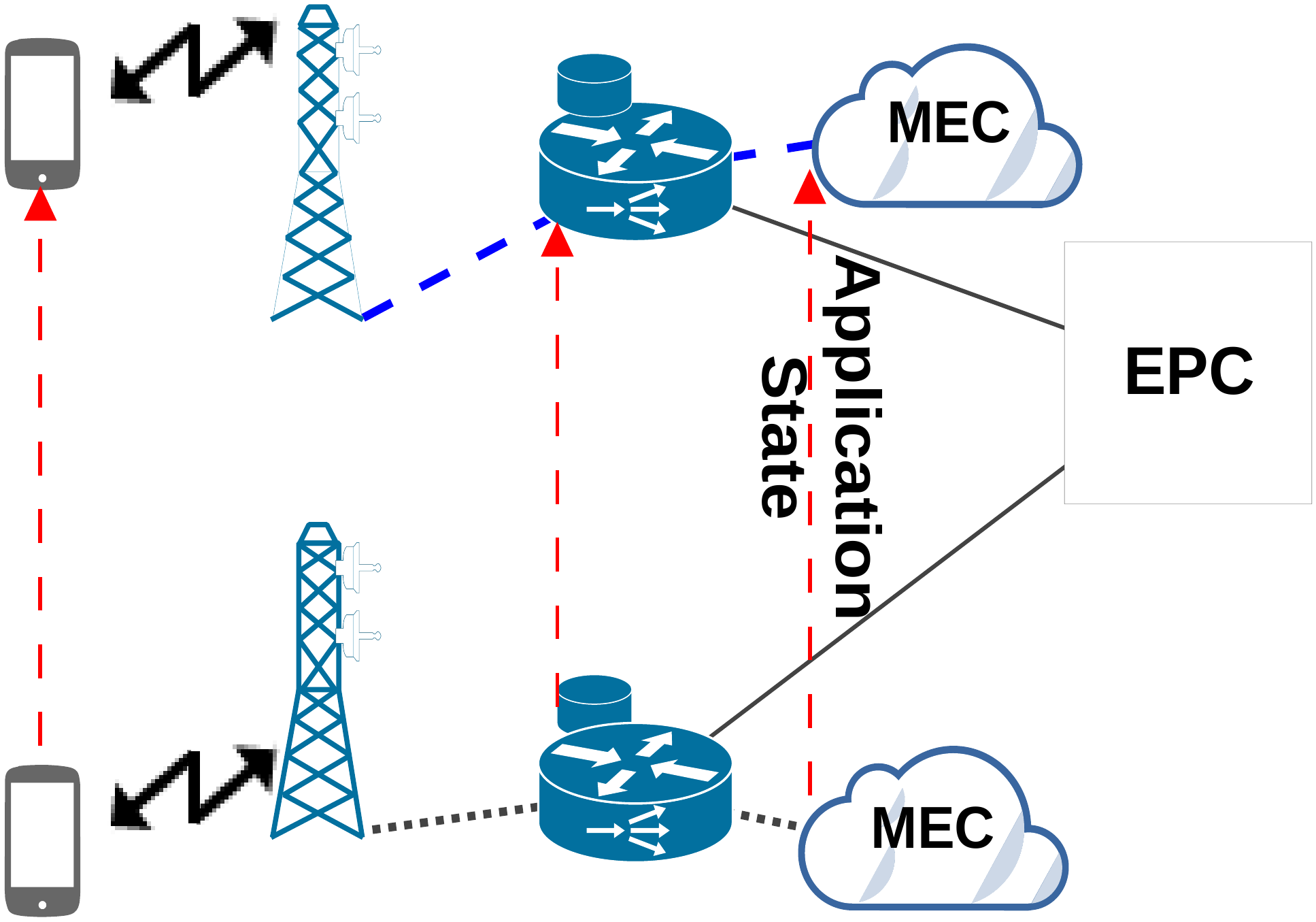}
		\label{migrationScenario3}
	\end{minipage}
	\caption{UE handover may or may not involve network and application state migration depending on the location of old and new eNBs.}
	\label{fig:ho:scenarios}
\end{figure*}

MEGW's S1AP processor behaves differently compared to our previous work as it does not write the reconstructed TEID pairs to UE mapping on the MEGW data plane immediately.
Instead, MEGW data plane clones never-seen-before upstream edge packets to the S1AP processor which in turn update the \emph{downstream} table of the EGW with the proper downstream TEID with regard to the observed upstream packet.
The reduced control plane traffic for each UE attach, or handover improves the scalability of MEGW as it reduces the number of rules that the controller installs.
Moreover, the new mechanism enables MEGW to support multiple bearers for each UE to edge services as individual connections' source and destination L4 port is mapped to the downstream TEID in the GTP context table of MEGW.

\subsection{X2 Handovers}
Figure~\ref{fig:ho:scenarios} depicts the possible scenarios that may occur during a handover.
As shown in the handover time line (Figure~\ref{fig:LTE:HO}), during step 3 S1AP processor observes the new and old eNB addresses and determines which of the three handover scenarios is in progress.

In the first scenario, the two eNBs are connected to the same MEGW.
Therefore, application state migration is not required in this case.
MEGW merely needs to update the UE's TEID pairs once the handover is complete.
Once the S1AP processor observes the end marker packet (step 5) it removes all of the occurrences of the old UE to TEID mapping and releases the old mappings.
Therefore, after this step and before the next step the MEGW drops downstream edge packets and clone upstream edge packets to the S1AP processor.
This behavior is consistent with the silent period of LTE standard reviewed in \S\ref{subsection:background:handover}.
At step 8 the S1AP processor observes the new TEID pairs, reconstructs the new UE to TEID pairs mapping, and installs downstream rules for newly cloned packets to the edge service using the new mappings.

In the second scenario, old and new eNBs are connected to different MEGWs; however, application state migration is not required because the involved MEGWs are within the same region.
Consistent with the previous case, the old MEGW detects the handover scenario at step 3 of the handover timeline.
The old MEGW observes step 5 of the handover and releases the old TEID pairs from the data plane, thus initiating the silent period.
The new MEGW only observes step 8 and constructs the TEID pairs' mappings for the UE.
The new eNB receives the path switch request acknowledgment of step 8 after the new MEGW, and when it starts to forward packets to the new MEGW, the GTP context is already there and the silent period will be finished.

In the third scenario, the two MEGWs belong to different regions, as such the application state handover will be triggered.
The networking state is handled just like the previous scenario.
The only difference is that the old MEGW will notify the application that migration is required for the involved UE to the new location.
Step 5 of the handover timeline (which is also the start of the silent period) is when the old MEGW notifies the application.
We schedule the application state migration at the same time as the silent period in order to minimize the service downtime.
Once the application state is migrated to the new MEC cluster, it will respond to the upstream edge requests and concludes the silent period.

The duration of the silent period is prolonged in the last case as application state migration involves transferring large volumes of data and it may also include VM/application container boot ups.
In the first two cases, however, the silent period lasts precisely the same duration as native LTE silent period during which SGW does not send any downstream packets.

\section{Evaluation}\label{section:results}

\begin{figure}[t]
    \centering
    \includegraphics[width=0.475\textwidth]{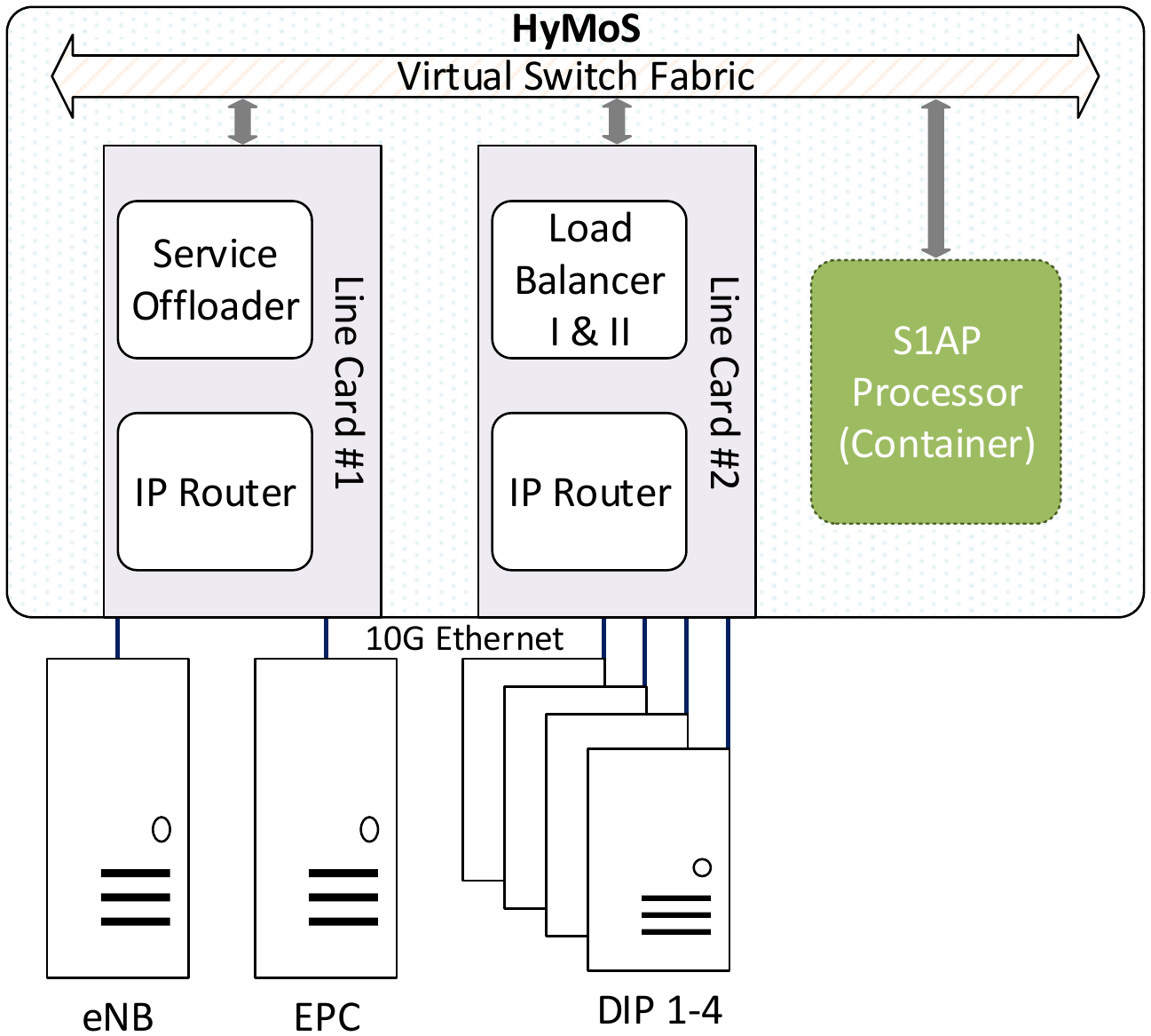}
    \caption{MEGW implemented using HyMoS with two Netronome NFP4000 line cards.}
    \label{fig:hymosModel}
\end{figure}

\subsection{HyMoS Testbed}

As shown in Figure~\ref{fig:hymosModel}, we have implemented the MEGW using our previously prototyped HyMoS~\cite{aghdai2017design}.
Our implementation uses two Netronome NFP 4000 line cards.
The first card implements the service offloader and IP router modules and is connected to two servers running OAI eNB and EPC via 10G Ethernet.
The second card implements load balancer I, II modules as well on top of an IP router that connects to four DIPs via 10G Ethernet.
The S1AP processor is implemented as an application container running in HyMoS and connected to the virtual fabric with a virtual network interface.
HyMoS virtualized fabric is a DPDK~\cite{intel2014data} application that routes packets between the line cards and connected virtual interfaces at high speed.

Using this testbed, we have verified the end-to-end operation of unmodified UE and LTE components with the addition of MEC enabled by MEGW.
Our proof-of-concept implementation works end-to-end.

Using OAISIM, we create several virtual Ethernet interfaces that pass the packets through the LTE protocol stack.
We have verfieid the Internet Traffic, MEC Traffic and S1AP traffic (both the cloned and the original ones) data paths.
The S1AP processor successfully listens to InitialContextSetupRequest/Response messages and installs UE to TEID mapping rules at the service offloader automatically.
The UE randomly reaches one of the DIPs if it sends a packet with the destination IP address of the VIP.
However, using this architecture we could not evaluate EGW performance as OAISIM and OAI protocol stack quickly becomes the bottleneck, limiting the throughput to less than 50Mbps.

\begin{figure*}[t]
    \centering
    \begin{minipage}[t]{0.245\textwidth}
        \documentclass[usenames,dvipsnames]{standalone}
\usepackage{pgfplots}
\usetikzlibrary{patterns}
\pgfplotsset{compat=newest}
\pgfplotsset{every tick label/.append style={font=\footnotesize}}

\begin{document}
    \begin{tikzpicture}
        \tikzstyle{every node}=[font=\Large]
        \begin{axis}[
                width  = 8cm,
                xlabel={Location of service},
                ylabel={End-to-end delay ($ms$)},
                height = 8cm,
                major x tick style = transparent,
                ybar=2*\pgflinewidth,
                bar width=9pt,
                ymajorgrids = true,
                symbolic x coords={MEC, PGW},
                xtick={data},
                scaled y ticks = false,
                enlarge x limits=0.50,
                ymin=0,
                ymax=300,
                legend cell align=left,
                legend columns=5,
                legend style={at={(0.5,1.045)},anchor=north},
            ]

            \addplot[style={fill=black},error bars/.cd, y dir=both, y explicit, error bar style=red]
            coordinates {
                (PGW, 267) += (0,0) -= (0,0)
                (MEC, 129) += (0,0) -= (0,0)
            };

        \end{axis}
\end{tikzpicture}
\end{document}
        \caption{End-to-end latency of OAI testbed.}
        \label{fig:e2eLatency}
    \end{minipage}
    \begin{minipage}[t]{0.36\textwidth}
        \documentclass[usenames,dvipsnames]{standalone}
\usepackage{pgfplots}
\pgfplotsset{compat=newest}
\pgfplotsset{every tick label/.append style={font=\footnotesize}}

\begin{document}

\begin{tikzpicture}
    \tikzstyle{every node}=[font=\Large]
	\begin{axis}[
		height=8cm,
		width=12cm,
		grid=major,
		xlabel={Packet Size (Bytes)},
		ylabel={Rate (Mpps)},
		xmin=0, xmax=1500,
                xtick={128,256,512,1024,1500},
		scatter/classes={%
		    a={mark=triangle*,red},%
		    b={mark=square*,blue},%
		    c={mark=o,draw=black}}]

        \addplot[domain=128:1500,mark=None,samples=1000] {1250/(x+20)};
	\addlegendentry{10G Line Rate}

	\addplot[scatter,only marks,mark options={scale=2},%
		scatter src=explicit symbolic]%
	table[meta=label] {
x     y      label
128	    8.30490    a
256     4.52828    a
512	    2.34925    a
1024	1.19719    a
1500	0.82228    a
	};
	]

	\addlegendentry{MEGW Rate}

	\end{axis}
\end{tikzpicture}

\end{document}
        \caption{MEGW throughput.}
        \label{fig:megwTp}
    \end{minipage}
    \begin{minipage}[t]{0.36\textwidth}
        \documentclass[usenames,dvipsnames]{standalone}
\usepackage{pgfplots}
\pgfplotsset{compat=newest}
\pgfplotsset{every tick label/.append style={font=\footnotesize}}

\begin{document}

\begin{tikzpicture}
    \tikzstyle{every node}=[font=\Large]
	\begin{axis}[
		height=8cm,
		width=12cm,
		grid=major,
		xlabel={Packet Size (Bytes)},
		ylabel={MEGW Latency ($\mu s$)},
		xmin=0, xmax=1500,
		ymin=0, ymax=180,
		xtick=data,
	]
		

	\addplot coordinates {
		(128, 155.3)
		(256, 142.2)
		(512, 148.9)
		(1024, 137.6)
		(1500, 133.6)
	};
	\end{axis}
\end{tikzpicture}

\end{document}
        \caption{Upper bound of MEGW latency.}
        \label{fig:megwLatency}
    \end{minipage}
    \label{fig:res:tb}
\end{figure*}

We have compared the end-to-end latency of a an application when it is installed at PGW to end-to-end latency of the same application installed on DIPs at MEC.
Results are shown in Figure~\ref{fig:e2eLatency}; the latency of the service at the edge is roughly half of that of the service installed at PGW.
Further analysis of the results showed that the OAI protocol stack at eNB and EPC is the main contributer to the measured delay.
Therefore, to better evaluate MEGW performance, we have installed a software packet generator instead of the eNB of Figure~\ref{fig:hymosModel}.
In the second test, the software packet generator sends GTP-U packets at variable size destined to VIP, and DIPs simply echo the received packet.
All packets have the same UE source IP address, and we have added a static rule to the service offloader with a hypothetical TEID for the software packet generator's UE.
First, we measure the throughput of MEGW and verify that it works at line rate with different packet sizes; measured packet processing rates are shown in Figure~\ref{fig:megwTp}.
Next, we measure the round trip time.
Packet traverse through MEGW twice in this scenario. The upper bound of processing time of MEGW is at most half as much as the measured round trip times.
The upper bound of MEGW latency for various packet sizes is shown in Figure~\ref{fig:megwLatency}.
The measured upper bound of latency is at most $150\mu s$ for the NPU-based P4 line cards.
While the measured delay is acceptable for most applications, it is possible to reduce the processing times by at least one or two orders of magnitude by using FPGA- or ASIC-based P4 line cards in HyMoS.

\subsection{Simulation Results}

\begin{figure}[t]
    \centering
    \includegraphics[width=0.475\textwidth]{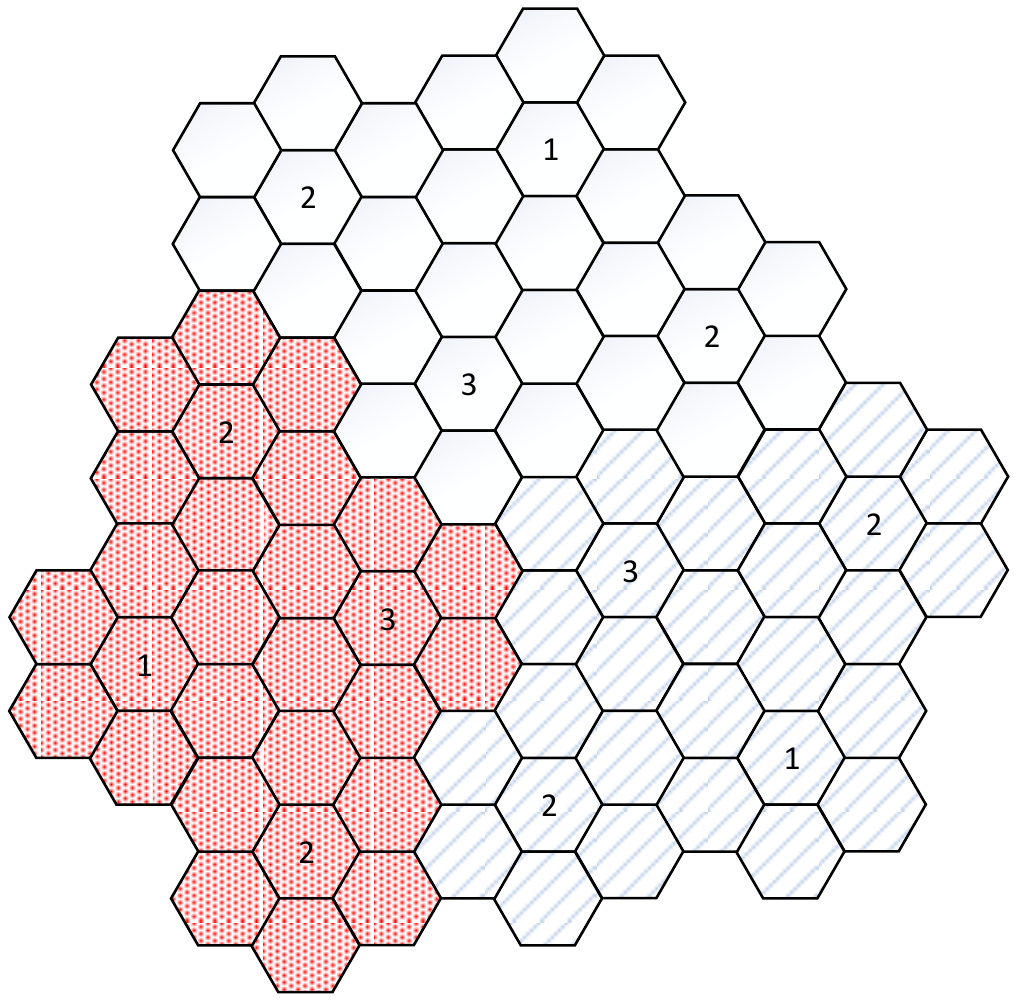}
    \caption{Topology of mobile network used in flow-level simulations.}
    \label{fig:simulationArch}
\end{figure}

We have used a flow-level simulation to evaluate the performance of the proposed 2-stage traffic steering method.
Figure~\ref{fig:simulationArch} illustrates the topology we used in our simulation.
Each hexagon represents a cell in the cellular network with a dedicated eNB.
Some cells are marked with a number; those cells have an MEC cluster.
The numbers represent the relative capacity of MEC cluster.
We assume that each MEC serves its cell as well as its 6 surrounding cells.
Finally, cells filled with the same pattern belong to the same region.
Our simulation includes three regions each containing four MECs.
We start the simulation with $500 * Capacity_MEC$ users connected to each MEC.
For example, the MEC that is marked by 2 and its 6 surronding cells are intialized with 1000 mobile phones randomly distributed within them.
We further assume that each mobile user consumes similar amount of resources in MEC.
Our simulation each minute migrates a number of randomly chosen users to a randomly chosen neighbor cell and measures two metrics: number of application state migrations and min-max ratio of MEC utilizations.
The rate at which users migrate among cells is variable.
We have replicated the measurements are 20 times for each data point.

\begin{figure*}[t]
    \centering
    \begin{minipage}[t]{0.45\textwidth}
        \documentclass[usenames,dvipsnames]{standalone}
\usepackage{pgfplots}
\pgfplotsset{compat=newest}
\pgfplotsset{every tick label/.append style={font=\footnotesize}}

\begin{document}

\begin{tikzpicture}
	\begin{axis}[
		height=6cm,
		width=10cm,
		grid=major,
        xlabel={Migrations per Second},
        ylabel={Application Migrations},
                xtick={10, 20, 50, 100, 200},
		scatter/classes={%
		    a={mark=square,blue},%
		    b={mark=triangle*,red},%
            c={mark=pentagon*,draw=black}},
            legend columns=2,
            legend style={at={(0.5,1.3)},anchor=north},
        ]

        \addplot[scatter,only marks,%
            scatter src=explicit symbolic]%
            table[meta=label] {
                x     y      label
                10	66.1      a
                20	131.05    a
                50	334.55    a
                100	654.05    a
                200	1301.5    a
            };
        ]
        \addlegendentry{With Regions}

        \addplot[scatter,only marks,%
            scatter src=explicit symbolic]%
            table[meta=label] {
                x     y      label
                10	210.8     b
                20	421       b
                50	1061      b
                100	2117.9    b
                200	4282.25   b
            };
        ]
        \addlegendentry{Without Regions}

	\end{axis}
\end{tikzpicture}

\end{document}
        \caption{Application migrations are reduced with 2-stage traffic steering.}
        \label{fig:simMigrations}
    \end{minipage}
    \begin{minipage}[t]{0.45\textwidth}
        \documentclass[usenames,dvipsnames]{standalone}
\usepackage{pgfplots}
\pgfplotsset{compat=newest}
\pgfplotsset{every tick label/.append style={font=\footnotesize}}

\begin{document}

\begin{tikzpicture}
	\begin{axis}[
		height=6cm,
		width=10cm,
		grid=major,
        xlabel={Migrations per Second},
        ylabel={Min-max Ratio},
                xtick={10, 20, 50, 100, 200},
		scatter/classes={%
		    a={mark=square,blue},%
		    b={mark=triangle*,red},%
            c={mark=pentagon*,draw=black}},
            legend columns=2,
            legend style={at={(0.5,1.3)},anchor=north},
        ]

        \addplot[scatter,only marks,%
            scatter src=explicit symbolic]%
            table[meta=label] {
                x     y      label
                10	0.997732060105216   a
                20	0.99659744728975    a
                50	0.995290868742117   a
                100	0.991021599853888   a
                200	0.985523928618199   a
            };
        ]
        \addlegendentry{With Regions}

        \addplot[scatter,only marks,%
            scatter src=explicit symbolic]%
            table[meta=label] {
                x     y      label
                10	0.971872127567869   b
                20	0.959642813918322   b
                50	0.912767753970052   b
                100	0.840514858159847   b
                200	0.757862358581125   b
            };
        ]
        \addlegendentry{Without Regions}

	\end{axis}
\end{tikzpicture}

\end{document}
        \caption{Min-max fairness is achieved with 2-stage traffic steering.}
        \label{fig:simFairness}
    \end{minipage}
\end{figure*}

We compare the measured metrics in two different scenarios:

\begin{enumerate}
    \item With regions: Where MEGWs use our proposed two-stage traffic steering.
        In this case if the used moves in the same region, application state will not migrate.
    \item Without regions: Where each MEC only serves its own cell and the 6 cells that surround it.
        In this case application state does not migrate if users move to a cell that is served by the same MEC.
        However, if users move to a cell that is served by a different MEC, application state migration will be triggered.
\end{enumerate}

Figure~\ref{fig:simMigrations} compares the number of application migrations with variable user migrations with and without the two-stage traffic steering in regions.
As expected, the two-stage algorithm with regions shows a linear improvement over the base algorithm.
With the two-stage region algorithm the number of application migrations remains less than 30\% of the base algorithms in all of our measurements.

Next we evaluate how fair each algorithm is.
We measure the average min-max ratio among MECs, i.e., the ratio between utilization of least-loaded MEC is devided to the utilization of highest-loaded MEC.
At the start of our simulation this ratio is 1 as we distribute the users in proportion to MECs' relative capacity.
However, as users move the min-max ratio degrades.
The min-max ratio directly impacts the quality of service.
Ideal case is where all users are assigned to MEC that have the same resource utilization (i.e., min-max ratio = 1).
When some MECs have higher utilization, their users will experience a larger delay, while users connected to MECs with lower utilization receive the service with smaller delay.
As such a lower min-max ratio indicates degraded quality of service for some users.
Figure~\ref{fig:simFairness} compares the average measured min-max ratios for the two-stage region traffic steering to that of the baseline.
Our measurements show that without using the two-stage algorithm on regions, the min-max ratio rapidly degrades as more users start to move in the network.
However, with the two-stage algorithm on regions, the min-max ratio degrades ever so slightly that it is not felt by the users.
This measurement also confirms our intuition in designing the two-stage algorithm.
Our proposed algorithm effectively makes a large virtual MEC cluster in each region out of small physical MECs.
Larger resource pools in the virtual MEC cluster of each region makes the utilization of the pool less sensitive to removal or addition of individual users.
In other words, larger size of the virtual clusters enable it to absorb the random behavior of users and achieve phenomenal min-max fairness.

\section{Related Works}\label{section:related}

Mobile edge computing~\cite{hu2015mobile} is a topic that is receiving much attention along with the standardization of the next generation of wireless networks.
Li et al. (2018)~\cite{li2018mobile} propose a softwarized middleware to intercept edge connections and route them toward MEC.
However, their solution neither supports mobility nor does it work if UEs use multiple bearer as they assume a one-to-one mapping between UE IP addresses and assigned TEIDs.
Ouyang et al. (2018)~\cite{ouyang2018follow} propose an intelligent service placement mechanism for edge services, however, they do not discuss how such mechanisms can be implemented in production networks.
Ye et al. (2016)~\cite{ye2016mobility} propose peer-to-peer communication among  UEs rather than going through eNBs.
However, such solutions require modifications in scheduling of radio access channels and only work for stationary UEs; as such mobility is not supported by these solutions.
Zhou et al. (2018)~\cite{zhou2018mec} explore a specific application of MEC as used in V2X systems.
It proposes a means to enable mobility for V2X servers as MEC applications; however, the solution is costly as it moves the application state for each handover.

There are many industry solutions for pushing compute and storage resources closer to mobile users.
M-CORD~\cite{M-CORD} and Magma~\cite{MAGMA} may be deployed in an MEC scenario.
M-CORD, however, requires multiple bearers and IP anchors for potentially different services.
The added complexity creates extra load in the mobile control plane as opposed to our approach that is completely backward compatible with the existing LTE and does not create extra load on its control plane.
MAGMA on the other hand, pushes the EPC to the radio access network and proposes Access Gateways that include their own softwarized EPC and are deployed in close proximity of eNBs.
While MAGMA breaks the vendor lock for EPC deployment and is very cheap to deploy, it does not support mobility;
if UEs move between eNBs connected to different access gateways, they would need to detach from old EPC and attach to the new EPC, effectively breaking all user connections.
Other industry solutions include Netflix' open connect appliance (OCA)~\cite{netflixOCA} and points of Presence (PoPs) deployed by CDNs such as Akamai~\cite{nygren2010akamai}.
However, these solutions are usually deployed at IP exchanges or PGW and are not close to the radio edge of mot pushhhhhbile networks.

Our work also relates to transport-layer load balancing solutions.
Silkroad~\cite{miao2017silkroad}, Ananta~\cite{patel2013ananta}, and Duet~\cite{gandhi2015duet} are examples of L4 load balancers that maintain connection affinity by means of assigning new connections using ECMP and tracking existing connections using a connection table at hardware, software, or combination of both, respectively.
Maglev~\cite{eisenbud2016maglev} uses a novel consistent hashing algorithm to distribute new flows among DIPs more uniformly while keeping track of existing connections in a connection table at software.
Faild~\cite{araujo2018balancing} and Beamer~\cite{olteanu2018stateless} use two-stage consistent hashing to implement stateless load balancing at the network; however, both solutions require cooperation in the form of rerouting of some connections from modified DIPs in certain circumstances to maintain connection affinity.
Spotlight~\cite{aghdai2018spotlight} introduces DIP utilization-aware adaptive weighted flow dispatching to distribute new connections uniformly while relying on a connection table to guarantee connection affinity.

\section{Conclusion}\label{section:conclusion}

We have designed, implemented, and verified the end-to-end operation of a softwarized MEC enabler in MEGW.
We show that by moving the mission-critical services to the radio edge of the network the ultra-low latency promise of 5G is achievable.
Furthermore, our solution does not require any modification to the existing components of networks, LTE protocol stack, and the applications that run on top of it.
MEGW is a practical solution to enable mobility at the edge-cloud.
Mobile edge-cloud is achieved by transparently listening to the LTE control plane without termination of its connection in addition to the usage of two-stage traffic steering mechanism that treats groups of small MECs as a single large virtual cluster in regions.
Our simulation results as well as testbed implementation show that the two-stage traffic steering not only decreases the number of application migrations, but also simplifies the service provisioning for MEC,

\bibliographystyle{ieeetr}
\bibliography{references}

\begin{thebibliography}{10}

\bibitem{5garch}
{\relax ETSI}, ``{\relax Technical Specification 123.501 System Architecture
  for the 5G System},'' {\em Release 15.2.0}, 2018.

\bibitem{zheng2015heterogeneous}
K.~Zheng, Q.~Zheng, P.~Chatzimisios, W.~Xiang, and Y.~Zhou, ``Heterogeneous
  vehicular networking: A survey on architecture, challenges, and solutions,''
  {\em IEEE communications surveys \& tutorials}, vol.~17, no.~4,
  pp.~2377--2396, 2015.

\bibitem{hu2015mobile}
Y.~C. Hu, M.~Patel, D.~Sabella, N.~Sprecher, and V.~Young, ``Mobile edge
  computing—a key technology towards 5g,'' {\em ETSI white paper}, vol.~11,
  no.~11, pp.~1--16, 2015.

\bibitem{chen2016efficient}
X.~Chen, L.~Jiao, W.~Li, and X.~Fu, ``Efficient multi-user computation
  offloading for mobile-edge cloud computing,'' {\em IEEE/ACM Transactions on
  Networking}, vol.~24, no.~5, pp.~2795--2808, 2016.

\bibitem{networkArch}
{\relax 3GPP}, ``{\relax Technical Specification 23.002 Network
  architecture},'' {\em Release 1999}, 2015.

\bibitem{li2018mobile}
C.~Li, H.~Liu, P.~Huang, H.~Chien, G.~Tu, P.~Hong, and Y.~Lin, ``Mobile edge
  computing platform deployment in 4{G} {LTE} networks: {A} middlebox
  approach,'' in {\em {USENIX} Workshop on Hot Topics in Edge Computing,
  HotEdge 2018, Boston, MA, July 10, 2018.}, 2018.

\bibitem{M-CORD}
{\relax Open Networking Foundation}, ``{\relax M-CORD Open Source Reference
  Solution for 5G Mobile Wireless Networks}.''
  \url{"https://www.opennetworking.org/m-cord/"}, accessed March 31, 2019.

\bibitem{MAGMA}
{\relax Facebook Connectivity}, ``{\relax MAGMA}.''
  \url{"https://github.com/facebookincubator/magma"}, accessed March 31, 2019.

\bibitem{aghdai2018transparent}
A.~Aghdai, M.~Huang, D.~Dai, Y.~Xu, and J.~Chao, ``Transparent edge gateway for
  mobile networks,'' in {\em 2018 IEEE 26th International Conference on Network
  Protocols (ICNP)}, pp.~412--417, IEEE, Sept 2018.

\bibitem{nfp4000}
Netronome, ``Nfp-4000 intelligent ethernet controller family.''
  \url{https://www.netronome.com/}.

\bibitem{nikaein2014openairinterface}
N.~Nikaein, M.~K. Marina, S.~Manickam, A.~Dawson, R.~Knopp, and C.~Bonnet,
  ``Openairinterface: A flexible platform for 5g research,'' {\em ACM SIGCOMM
  Computer Communication Review}, vol.~44, no.~5, pp.~33--38, 2014.

\bibitem{eutran}
{\relax 3GPP}, ``{\relax Technical Specification 36.401 Evolved Universal
  Terrestrial Radio Access Network},'' {\em Release 8}, 2015.

\bibitem{machen2018live}
A.~Machen, S.~Wang, K.~K. Leung, B.~Ko, and T.~Salonidis, ``Live service
  migration in mobile edge clouds,'' {\em {IEEE} Wireless Commun.}, vol.~25,
  no.~1, pp.~140--147, 2018.

\bibitem{gtpu}
{\relax 3GPP}, ``{\relax Technical Specification 29.281 GPRS Tunneling Protocol
  User Plane (GTPv1-U)},'' {\em Release 15}, 2018.

\bibitem{s1ap}
{\relax 3GPP}, ``{\relax Technical Specification 36.413 S1 Application Protocol
  (S1AP)},'' {\em Release 8}, 2015.

\bibitem{steedman1993abstract}
D.~Steedman, {\em Abstract syntax notation one (ASN. 1): the tutorial and
  reference}.
\newblock Technology appraisals, 1993.

\bibitem{NAS}
{\relax 3GPP}, ``{\relax Technical Specification 24.301 Non-Access-Stratum
  (NAS) protocol for Evolved Packet System (EPS)},'' {\em Release 8}, 2015.

\bibitem{miao2017silkroad}
R.~Miao, H.~Zeng, C.~Kim, J.~Lee, and M.~Yu, ``Silkroad: Making stateful
  layer-4 load balancing fast and cheap using switching asics,'' in {\em
  Proceedings of the Conference of the {ACM} Special Interest Group on Data
  Communication, {SIGCOMM} 2017, Los Angeles, CA, USA, August 21-25, 2017},
  pp.~15--28, 2017.

\bibitem{olteanu2018stateless}
V.~A. Olteanu, A.~Agache, A.~Voinescu, and C.~Raiciu, ``Stateless datacenter
  load-balancing with beamer,'' in {\em 15th {USENIX} Symposium on Networked
  Systems Design and Implementation, {NSDI} 2018, Renton, WA, USA, April 9-11,
  2018}, pp.~125--139, 2018.

\bibitem{aghdai2018spotlight}
A.~Aghdai, C.~Chu, Y.~Xu, D.~H. Dai, J.~Xu, and H.~J. Chao, ``Spotlight:
  Scalable transport layer load balancing for data center networks,'' {\em
  CoRR}, vol.~abs/1806.08455, 2018.

\bibitem{lteho1}
{\relax 3GPP}, ``{\relax Technical Specification 36.300 E-UTRAN overall
  description},'' {\em Release 11}, 2013.

\bibitem{lteho2}
{\relax 3GPP}, ``{\relax Technical Specification 23.401 GPRS enhancements for
  E-UTRAN access},'' {\em Release 11}, 2013.

\bibitem{att2017aec}
{\relax AT\&T inc.}, ``A{T}\&{T} {AEC} {(AT\&T Edge Cloud)} architecture white
  paper,'' 2017.

\bibitem{att2016ecomp}
{\relax AT\&T inc.}, ``A{T}\&{T} {ECOMP} {(Enhanced Control, Orchestration,
  Management, and Policy)} architecture white paper,'' 2016.

\bibitem{aghdai2017design}
A.~Aghdai, Y.~Xu, and H.~J. Chao, ``Design of a hybrid modular switch,'' in
  {\em Network Function Virtualization and Software Defined Networks (NFV-SDN),
  2017 IEEE Conference on}, p.~6, IEEE, 2017.

\bibitem{intel2014data}
Intel, ``Data plane development kit,'' 2014.

\bibitem{ouyang2018follow}
T.~Ouyang, Z.~Zhou, and X.~Chen, ``Follow me at the edge: Mobility-aware
  dynamic service placement for mobile edge computing,'' {\em IEEE Journal on
  Selected Areas in Communications}, vol.~36, no.~10, pp.~2333--2345, 2018.

\bibitem{ye2016mobility}
C.~Ye, P.~Wang, C.~Wang, and F.~Liu, ``Mobility management for lte-based
  heterogeneous vehicular network in v2x scenario,'' in {\em 2016 2nd IEEE
  International Conference on Computer and Communications (ICCC)},
  pp.~2203--2207, IEEE, 2016.

\bibitem{zhou2018mec}
S.~Zhou, P.~P. Netalkar, Y.~Chang, Y.~Xu, and H.~J. Chao, ``The {MEC}-based
  architecture design for low-latency and fast hand-off vehicular networking,''
  in {\em {\relax Vehicular Technology Conference (VTC), 2018 IEEE Conference
  on}}, p.~7, IEEE, 2018.

\bibitem{netflixOCA}
{\relax Netflix}, ``{\relax Open Connect Appliance}.''
  \url{"https://openconnect.netflix.com"}, accessed June 20, 2018.

\bibitem{nygren2010akamai}
E.~Nygren, R.~K. Sitaraman, and J.~Sun, ``The akamai network: a platform for
  high-performance internet applications,'' {\em ACM SIGOPS Operating Systems
  Review}, vol.~44, no.~3, pp.~2--19, 2010.

\bibitem{patel2013ananta}
P.~Patel, D.~Bansal, L.~Yuan, A.~Murthy, A.~G. Greenberg, D.~A. Maltz, R.~Kern,
  H.~Kumar, M.~Zikos, H.~Wu, C.~Kim, and N.~Karri, ``Ananta: cloud scale load
  balancing,'' in {\em {ACM} {SIGCOMM} 2013 Conference, SIGCOMM'13, Hong Kong,
  China, August 12-16, 2013}, pp.~207--218, 2013.

\bibitem{gandhi2015duet}
R.~Gandhi, H.~H. Liu, Y.~C. Hu, G.~Lu, J.~Padhye, L.~Yuan, and M.~Zhang,
  ``Duet: Cloud scale load balancing with hardware and software,'' {\em ACM
  SIGCOMM Computer Communication Review}, vol.~44, no.~4, pp.~27--38, 2015.

\bibitem{eisenbud2016maglev}
D.~E. Eisenbud, C.~Yi, C.~Contavalli, C.~Smith, R.~Kononov,
  E.~Mann{-}Hielscher, A.~Cilingiroglu, B.~Cheyney, W.~Shang, and J.~D. Hosein,
  ``Maglev: {A} fast and reliable software network load balancer,'' in {\em
  13th {USENIX} Symposium on Networked Systems Design and Implementation,
  {NSDI} 2016, Santa Clara, CA, USA, March 16-18, 2016}, pp.~523--535, 2016.

\bibitem{araujo2018balancing}
J.~T. Ara{\'{u}}jo, L.~Saino, L.~Buytenhek, and R.~Landa, ``Balancing on the
  edge: Transport affinity without network state,'' in {\em 15th {USENIX}
  Symposium on Networked Systems Design and Implementation, {NSDI} 2018,
  Renton, WA, USA, April 9-11, 2018}, pp.~111--124, 2018.

\end{thebibliography}

\end{document}